\documentclass[smallextended]{svjour3}       
\smartqed  
\usepackage{graphicx}
\usepackage{xcolor}
\newcommand\bef{\begin{figure}}
\newcommand\eef[1]{\label{fg:#1}\end{figure}}
\newcommand\beq{\begin{equation}}
\newcommand\eeq[1]{\label{#1}\end{equation}}
\newcommand\beqa{\begin{eqnarray}}
\newcommand\eeqa[1]{\label{#1}\end{eqnarray}}

\newcommand\fgn[1]{Figure \ref{fg:#1}}
\newcommand\eqn[1]{eq.\ (\ref{#1})}

\newcommand{\etc}{{\sl etc.\/}}
\newcommand\ie{{\sl i.e.\/}}
\newcommand{\qcd}{QCD$_2$}
\begin{document}

\title{Fun with colours
}
\subtitle{The standard model with two colour QCD has radically different
 long distance physics}


\author{Saumen Datta,\\ Sourendu Gupta,\\ Rishi Sharma
 }


\institute{
              Department of Theoretical Physics, 
              Tata Institute of Fundamental Research,\\
              Homi Bhabha Road, Mumbai 400005, India.\\
              SD's \email{saumen@theory.tifr.res.in},\\
              SG's \email{sgupta@tifr.res.in},\\
              RS's \email{rishi@theory.tifr.res.in}           
}

\date{Received: date / Accepted: date}

\maketitle

\begin{abstract}
In our world the standard model of particle physics contains within it
the fairly intractable theory called QCD. A toy version with two colours
is often studied as a model confining and chiral symmetry breaking field
theory. Here we investigate the cascade of changes at various distance
scales if we make this change within the standard model. It is possible
to limit the changes at the hadronic scale. However, the minor changes
that occur actually cascade down to the far infrared, into nuclear and
atomic physics, and chemistry. Through this it also possibly affects the
evolution of stars and galaxies. We remark on this unexpected sensitivity
of the universe to physics at the scale of quarks.
\end{abstract}

\section{Introduction}\label{intro}

In our universe the standard model (SM) of particle physics \cite{weinberg}
contains strong interactions described by a gauge group SU($N_c$) with
$N_c=3$ and with six flavours of quarks in the fundamental representation
of the gauge group. This is quantum chromodynamics, QCD. The action has a
global chiral symmetry, and quarks are, additionally, in the fundamental
representation of this symmetry. The Glashow Iliopoulos Maiani (GIM)
mechanism organizes these quark flavours into three generations, each
of two flavours. The currents built out of the quark doublets and the
corresponding lepton doublets are coupled to the SU(2)$\times$U(1)
gauge group of the electroweak interactions.  A matching of the number
of generations is required by the GIM anomaly cancellation mechanism. The
Higgs mechanism does not depend on QCD\footnote{This can change if the
hierarchy of scales in the standard model is modified drastically
\cite{phase}, which we do not plan to do.}.

The electroweak part of the theory is reasonably well described in
a weak coupling expansion. but the strong interactions are not, in
general. In particular, the external states of the SM are
not quarks and leptons but leptons and hadrons. This absence of quarks
from the external states is said to be due to a property of QCD called
confinement, whose understanding remains incomplete in spite of a million
dollar prize for a proof. Since the strong interaction is resistant to a
weak coupling expansion, methods such as lattice field theory. the large
$N_c$ (number of colours) expansion, effective field theories (EFTs),
\etc, have been developed to deal with it.

Our current understanding is that confinement holds for all $N_c$ when
the number of light quark flavours is small enough. As a result, one
may study arbitrary $N_c$ for the insights it gives into the central
problem of the confinement of quarks inside hadrons, and the lack of
asymptotic quark states. We argue here that although these insights
and arguments are quite valid, a world with different values of $N_c$
can behave quite differently.

We take the case of $N_c=2$ as an example. This was once a popular
toy model in lattice studies of the strong interactions \cite{chiral}.
Not only does the smaller number of components of the gauge and quark
fields make it easier to work with, the fundamental representation of
SU(2) is self-contragradient.  As a result the integration measure in
the path integral at finite baryon density is real. This allows direct
numerical simulation at finite chemical potential, something which is
not possible for SU(3) colour. The realization that the phase diagram
and physics of {\qcd} is quite different from that of QCD has dampened
some of the enthusiasm.  However, these results have recently leaked
out into the wider world of nuclear physics \cite{leak}, and this is
where some caution needs to be exercised.

Here we explore the physics of {\qcd}, the theory of strong interactions
with $N_c=2$ for the gauge group, with $N_f$ quark flavours in the
fundamental of the colour gauge group, and an unchanged electroweak
sector, over many different length scales. This note is meant to serve
multiple purposes.  First, one can have fun constructing the physics
of a world with two colours, in a long tradition of speculative fiction
\cite{fiction}.  Second, it sharpens the questions one can (and cannot, yet)
answer about our world by examining this counterfactual universe. An
example is the answer to the question: given our knowledge of weak
interactions, what is the simplest low energy experiment that shows that
our SM has $N_c=3$ and not 2? One surprising answer is that
$N_c$ is an odd number since we know that Hydrogen atoms exist. Third,
an exploration like this may reveal technical niceties about the real
world which are worth exploring. Pushan Majumdar was involved with the
authors in an unfinished computation which led on from such speculation,
and we present this work as a discursive memorial to a friend. Finally,
our exploration illustrates a larger technical issue for the effective
field theory program, since the physics at the nuclear, atomic, molecular,
and longer length scales seems to know about changes at the length scale
of quarks. Through this exploration we probe how to interpret decoupling
theorems in quantum field theory.

\section{Particle physics}

We state the model before starting\footnote{We follow the metric and
other conventions of Weinberg \cite{weinberg}}. The lepton and electroweak
sector of the SM remains unchanged. We will introduce quark flavours in
pairs, so that the GIM mechanism continues to operate, and take $u$ and
$d$ to be light.  The SU(2) anomaly \cite{wanomaly} forces us to have two
or four generations of fermions \cite{shrock}.  Where the other flavours
are concerned, one may take a slightly heavier $s$, and significantly
heavier other quarks.  Apart from this rough hierarchy, we allow the
masses to be generic. When we consider the chiral limit, we take two
or three lightest flavours to vanish.  The substantive change is that
$N_c=2$, so the strong interactions are described by {\qcd} and not QCD.

\subsection{Quarks}

We have made a minimal change to the SM. The charged weak
current connecting quarks and leptons continues to allow decays such
as $s\to u\ell\overline\nu_\ell$ in \qcd, through the usual Cabibbo
Kobayashi Maskawa mixing of quark flavours.  This also requires the
charges of all up type and down type quarks to be equal, and the two
charges to differ by unity, \ie,
\beq
 Q_u-Q_d=1
\eeq{charge}
The Adler Bell Jackiw (ABJ) anomaly cancellation \cite{weinberg} also gives
the relation
\beq
   Q_u^2-Q_d^2=\frac1{N_c}.
\eeq{gim}
These two together give $Q_u=(1+N_c)/(2N_c)$ and $Q_d=(1-N_c)/(2N_c)$
\cite{shrock,abbas,chow}.  The $u$ is positively charged and the
$d$ is negative for all $N_c$.  For $N_c=2$ we find that $Q_u=3/4$
and $Q_d=-1/4$. It is interesting that in the large $N_c$ limit
$Q_u=-Q_d=1/2$.

We will assume that confinement holds in \qcd.  Then all meson and baryon
states are colour singlet. Mesons can be built from quark antiquark
pairs, exactly as in QCD. At large $N_c$ baryons are built using $N_c$
quarks \cite{witten}. So the baryon number of each quark is taken to be
$1/N_c$. It is natural to extend this counting to $N_c=2$. Then a baryon
in {\qcd} is a diquark state. The fact that in {\qcd} the baryon is a
boson turns out to be consequential.

The first difference between our world and this alternative occurs at
this point. All the baryons that one can build from the first generation
of quarks, \ie, the {\qcd} analogues of the nucleon, have fractional
charge. Confinement does not rule them out. In fact, the observational
absence of fractionally charged particles in our world is enough to rule
out all even values for $N_c$.  The argument is the following. Take a
baryon with $n$ u quarks and the remaining d quarks. Its charge is $n
Q_u+(N_c-n) Q_d=n-(N_c-1)/2$ \cite{abbas}. This is an integer only for
odd $N_c$.

\subsection{Hadrons}

A basic organizational tool in hadronic physics is the quark model. The
chiral symmetry of QCD, for $N_f$ flavours of massless quarks, is
described by a global symmetry group G($N_f$). It is spontaneously
broken to the flavour subgroup H($N_f$) by a flavour-singlet quark
condensate. As a result, the particle spectrum is classified by H($N_f$).
In particular, the Goldstone bosons all lie in the same representation
of H($N_f$). In QCD G($N_f$) is SU$_L$($N_f$) $\times$ SU$_R$($N_f$)
$\times$ U$_V$(1), where the two SU($N_f$) factors act on the left
and right handed components of the quarks, the vector flavour singlet
U$_V$(1) corresponds to the baryon number symmetry, and the quantum
theory has no axial U$_A$(1) symmetry. This is broken to H($N_f$),
which is U$_V$($N_f$).  The $N_f^2-1$ Goldstone bosons are pseudoscalar
mesons which all lie in the same representation of SU$_V$($N_f$). $N_f=3$
gives Gell-Mann's eightfold way.

When the gauge group of the strong interactions is SU(2), it turns
out that a Pauli-G\"ursey mechanism enlarges the global symmetry group
to SU($2N_f$) (see \cite{stern} for details). This fact has been used
for model building in the past \cite{peskin}.  The largest subgroup,
H($N_f$), which results from a spontaneous symmetry breaking through
the formation of a flavour singlet condensate is the unitary group
USp($N_f$) = SU($2N_f$) $\cup$ Sp($2N_f$). This contains the usual flavour
group U$_V$($N_f$) as a subgroup.  The number of Goldstone bosons is
$(N_f-1)(2N_f+1)$, and they all lie in one irrep of USp($N_f$). They are
$N_f^2-1$ pseudoscalar mesons and $N_f(N_f-1)$ baryons. The fact that they
would lie in different irreps of U$_V$($N_f$) is consistent with the fact
that USp($N_f$) is larger, and contains generators which connect them.

One can proceed to write down an effective field theory of the
interacting Goldstone bosons. In such a theory all operators which are
allowed by symmetry should be included. This means that operators which
change the baryon number could also appear. This is another difference
between this world and ours.

Forbidding baryon number changing currents from the symmetry broken
theory is equivalent to asking H($N_f$) to be U$_V$($N_f$). In this
case the number of Goldstone bosons is $3N_f^2-1$. This requires a much
larger irrep of SU($N_f$) than the adjoint which contains the pseudoscalar
mesons. One also sees that an irrep as large as this cannot be made either
of $\overline qq$ mesons or $qq$ baryons, since both these constructions
give irreps no larger than $N_f^2$. As a result, this kind of symmetry
breaking would also yield a particle physics very different from that
in the SM.

\bef
\begin{center}
\includegraphics[scale=0.5]{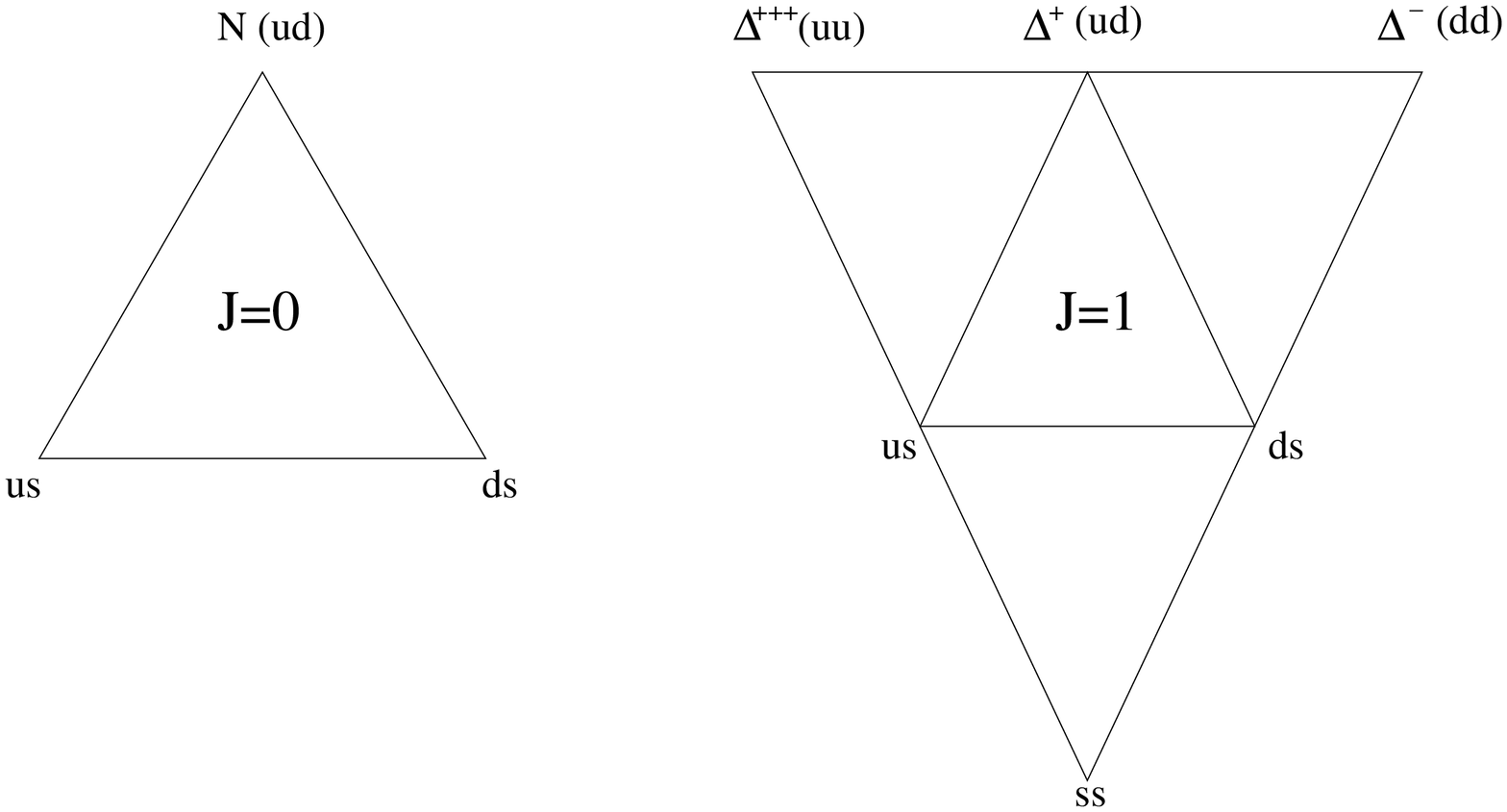}
\end{center}
\caption{The lowest multiplets of baryons in \qcd, with 3 flavours of quarks,
 when the chiral symmetry group obeys baryon number conservation. The
 non-strange baryons are given names which are used in the discussion of the
 nuclear physics of this world.}
\eef{baryon2}

The origin of these oddities is the Pauli-G\"ursey construction allowed
in {\qcd} \cite{stern}.  The fact that the fundamental representation
of SU(2) is self-contragradient allows one to construct the following
global transformation which commutes with the gauge symmetry
\beq
  \psi_f \to a\psi_f - b\psi_f^c, \quad{\rm where}\quad
    |a|^2+|b|^2=1,
\eeq{paulig}
with the notation that $\psi^c$ is the charge conjugate of the quark
field $\psi$, and $f$ is a flavour index. The corresponding current
allows us to violate baryon number conservation. If one wants a conserved
baryon number, then one can legislate that such transformations are not
allowed. The global symmetry group then remains what it was for $N_c=3$,
and this is broken to the usual flavour group, U$_V$($N_f$) by the
condensate.  The Goldstone bosons are then exactly the same pseudoscalar
mesons, which lie in an adjoint representation of SU($N_f$). The baryon
multiplets are distinct, and Weingarten's theorem \cite{weingarten}
tells us that all of them are massive.

The simplest way to impose baryon number conservation is to introduce it
as a constraint into the path integral of the theory. As usual, this can
be implemented in the form of a term in the Lagrangian that breaks the
larger SU($2N_f$) symmetry to the desired one explicitly. The extra term
is 
\beq
 L_B = {\cal B} \sum_{f=1}^{N_f}\overline\psi_f\gamma_0\psi_f
\eeq{baryon}
whose coefficient, $\cal B$, is analogous to a magnetic field which
breaks the SU($2N_f$) symmetry. We will call this a baryomagnetic field. A
side-effect of this term is to lift the degeneracy between the baryon and
the pion, and to make the former more massive \cite{kogut}. The scalar
baryon mass is proportional to $\cal B$ when this parameter is not too
large. 

For $N_f=2$ the Goldstone bosons are an isotriplet of pions, with the usual
charges. The scalar baryon is a isosinglet $ud$ combination, and has charge
1/2.  There is an isotriplet of vector baryons, with charges 3/2, 1/2,
and $-$1/2. The isotriplet of baryons is heavier than the isosinglet for
generic values of $\cal B$.

For $N_f=3$, the Goldstone bosons are the usual octet of pseudoscalar
bosons, the $\pi^0$, $\pi^\pm$, $K^0$, $\overline K^0$, $K^\pm$, and
$\eta$. The scalar baryons are in a $\overline 3$ of ud, us, and sd states
(charges 1/2, 1/2, and $-$1/2 respectively). There is also a sextet of
vector baryons (see \fgn{baryon2}).

One may legitimately ask whether non-quark model states such as glueballs
could be as light as (or lighter than) the baryons. The answer is hard to
give without a lattice computation. However, one could argue \cite{witten}
that when one changes $N_c$ and fixes the scale of QCD by an independent
scale such as the glueball mass, then the baryon mass is proportional
to $N_c$. For sufficiently large $N_c$ glueballs could become lighter
than baryons.  However, for $N_c=3$ lattice computations indicate that
the glueball mass is about 50\% more than the baryon mass. As a result,
baryon masses could be expected to be significantly smaller than the
glueball mass for $N_c=2$.  This counting could be suspect at $N_c=2$,
but the argument through the breaking of Pauli-G\"ursey symmetry also
leads us to believe this.

We close this part of the discussion with a brief remark on another
interesting issue. The chiral anomaly of QCD persists into \qcd.
The amplitude of the decay $\pi\to2\gamma$ involves the ABJ anomaly
term, and contains the coefficient $N_c(Q_u^2-Q_d^2)/f_\pi$, which
goes as $1/\sqrt N_c$, using \eqn{gim} and taking $f_\pi\propto\sqrt
N_c$ \cite{donoghue}.  The Wess-Zumino-Witten term gives the amplitude
for the reaction $K^+K^-\to\pi^+\pi^-\pi^0$ to be order $N_c$. Taking
into account the appropriate power of $f_\pi$, the amplitude for this
reaction is proportional to $1/N_c^{3/2}$. This power counting assumes
the dominance of planar graphs, and could be inaccurate in scaling from
$N_c=2$ to 3.

\subsection{Phase diagram}

The discussion of the phase diagram of {\qcd} in the literature is in the
context of the breaking of SU($2N_f$) symmetry. 
This has a conserved current which mixes
quark fields with their charge conjugates. 
As a result, the diquark
condensate and the chiral condensate become related. These studies are
not relevant to the version of {\qcd} we discuss.

Lattice studies of the phase diagram of the model we are interested
in would have to include the term in $\cal B$ in the action. In the
Euclidean action this would appear as an imaginary part to the baryon
chemical potential, $\mu$. It would have to be tuned to fix the nucleon
mass at a fixed lattice cutoff, and then left unchanged for a scan of the
real part of the chemical potential. In this set up the path integral of
{\qcd} at finite baryon chemical potential is no longer real. In spite of
the sign problem, working with an SU(2) gauge group takes about a third
of the CPU time that SU(3) takes. Since the strong interaction in this
model has physics very similar to that of QCD, a lattice study could be
worth setting up.

In particular, the universality argument, that phase diagrams are determined
by global symmetries, leads us to the conjecture that the Columbia plot of
{\qcd}, with explicitly broken Pauli-G\"ursey symmetry, is similar to that of QCD. Furthermore, as we remark later, the
strength and range of the NN interactions seem to be independently tunable.
Issues such as this would be interesting to explore. In the rest of this
note we will assume that the NN potential is attractive in \qcd, as it
is in QCD.

\section{Low energy physics}

\subsection{Nuclear physics}

Is there nuclear physics in this world? If the masses of the quarks are
similar to those in our world, then the only beta-stable non-strange
baryons are the isosinglet scalar $ud$ (which we will call the nucleon,
N), and the vector $uu$ (which we will call $\Delta^{+++}$). It is
interesting to note that once baryon number conservation is imposed
on the theory, and confinement holds, the lightest baryon must be
stable. The vector $ud$ ($\Delta^+$) can go to $\Delta^{+++}$ by a weak
decay, and the $dd$ ($\Delta^-$) can decay to N or $\Delta^+$ by the
weak interactions, and more rarely to $\Delta^{+++}$. Also possible
are the strong decays, $\Delta^{+++}\to N\pi^+$, $\Delta^+\to N\pi^0$,
and $\Delta^-\to N\pi^-$.  As a result, we expect the vector baryons
to play a minor role in the structure of nuclei, just as the spin 3/2
$\Delta$s do in our world \cite{delta}.  The chart of nuclei will be
simple; it will contain integer baryon number $B$ and charge $Z=B/2$.
Since the isosinglet N is the only component of nuclei, there can be no
isotopes. Also, since $N$ has half integer charge, it might be possible
to have nuclei which do not form neutral atoms.

Let us examine the question of nuclear physics in slightly more detail.
Our understanding of the long-range nuclear force in our world as due
to pion-nucleon interactions is related to QCD through baryon chiral
perturbation theory. The leading piece of this action has an Yukawa
interaction between pions and nucleons which is non-vanishing even when
the nucleon is at rest (see \cite{becher} for a review).  Since baryons
remain massive in \qcd, there will be non-vanishing interactions
between the bosonic nucleon and pions. Their origin is interesting enough
to merit some discussion.

When the baryomagnetic field $\cal B$ vanishes, the SU($2N_f$) symmetry
breaking gives us a set of Goldstone fields $\mathbf\phi$, which contain
both nucleons and pions. As a result, the interactions between them 
vanish in the static limit. Interaction terms such as $\mathbf\phi^4$
have coefficients which are linear in $\cal B$ and vanish in the chiral
limit. So the dimension-4 terms
which give rise to nucleon-nucleon forces are of the form 
\beq
   L_{NN\pi\pi} = \lambda \sum_{\alpha,a}
          N_\alpha^\dag N_\alpha\,\pi_a^\dag\pi_a
\eeq{nucleonforce}
where N is the scalar baryon field, $\pi$ the pseudoscalar meson field,
and $\alpha$ and $a$ are flavour indices taking values in the appropriate
multiplet. For $N_f=2$, N is isoscalar, so $\alpha$ is redundant, and
$a$ takes value between 1 and 3. $M$, the mass of N, is proportional
to $\cal B$, as is $\lambda$. So we can write $\lambda\propto M$. This
parameter controls the strength of the NN interaction. The range of the
interaction is given by the pion mass, $m$. Taking $m_q$ to be the light
quark mass, one has, as usual, $m\propto\sqrt{m_q}$. So in this version
of the strong interactions there are independent parameters to tune the
range and strength of the NN forces. The sign of $\lambda$ can be arranged
to give attractive interactions at long distances. 

At very short distances one should expect repulsion if stable tetraquark
(the {\qcd} analogue of dibaryon), triquark-antiquark (the analogue of
pentaquark) or other exotic baryon states are absent in the light quark
sector.  Then nuclei can form, and nuclear matter away from the chiral
limit is roughly incompressible, as in our world \footnote{Both the
intermediate distance regime, where multi-meson exchanges and the tower
of higher spin mesons have to be accounted for, and the chiral limit,
where intermediate states with infinite number of pions is not suppressed
by energy considerations, require a more detailed analysis. If nuclear
physics and bulk nuclear matter in {\qcd} were of greater interest,
these effects would have to be computed.}. It is noteworthy that in this
world the spectrum of the $B=2$ nucleus can be tuned by changing $M$
and $m$. One can have a situation with a complex nuclear spectrum for
$B=2$, or a single shallow bound state, as in our world, or the even a
situation where $B=2$ is unbound but $B=3$ state is a Borromean nucleus
\cite{borromean}. More generally, studies of pairwise interacting bosons
show \cite{efimov} that the $B=3$ and $B=4$ nuclei are likely to exist
in \qcd.

Since N is spinless, one may first think that the shell structure of
nuclei is irrelevant in \qcd, and all nuclei have spin zero.  However,
for large enough $B$, the Coulomb repulsion between the $N$s would
eventually make it favourable to populate a higher quantum state in the
effective NN potential. Consider the nuclear Hamiltonian
\beq
  H=\sum_{i=1}^B \left[\frac{p_i^2}{2M} + V(r_i)\right]
     + \frac{e^2}4\sum_{i>j}\frac1{|r_i-r_j|},
\eeq{ham}
where $M$ is the mass of the $N$, $e$ is the charge of the electron, and
$V$ is a short range potential which arises in the usual way through
the nuclear mean field. For simplicity choose the simplest shell
model potential, $V(r)=kr^2/2$. The single particle level spacing is
$\Omega=\sqrt{k/M}$. Taking the Coulomb energy term as a perturbation, one
sees that when $B(B-1)/2$ is large enough, the contribution of this term
may exceed $\Omega$. When this happens, it is energetically favourable
to start filling the next shell. As a result, there is a rudimentary
shell structure of the nucleus. Clearly, this argument is not specific
to the choice of $V(r)$.

There is an interesting piece of physics one can investigate here. The
mean field potential $V(r)$ arises from the NN potential. Since it is
short ranged, the number of bound states is finite. As a result, when $B$
is larger than some critical $B_d$, the Coulomb repulsion will become
so large that there will not be a bound state which accommodates all $B$
nucleons. This is the {\qcd} analogue of drip lines in our world. Since
the NN potential is tunable by changing $M$ and $m$, one may be able to
investigate the dependence of $B_d$ on $M$ and $m$. It is particularly
interesting to ask whether it is possible to have macroscopic charge
neutral objects which are not bound by gravity, but by the strong
interactions.

In our world one can describe the binding energies of nuclei in a liquid
drop model \cite{gamow}. This depends mainly on the existence of a short
range effective central potential, $V$ in \eqn{ham}. So it should continue
to work for nuclear physics built over \qcd. The volume term is due to the
mean-field averaged nuclear potential, $V(r)$ in \eqn{ham}, the surface
term is due to remnant two body nuclear forces which are not shown in
\eqn{ham}, and the origin of the Coulomb repulsive term is clear. The
asymmetry and pairing terms, which account for isotopes, are absent.
The binding energy per nucleon will therefore first increase slightly
with $B$ because of the remnant two body terms, and then decrease due
to the Coulomb term.

Is there radioactivity in this world? Nuclei will have excited states,
which could decay radiatively. Weak decays are ruled out, since N is
stable and isoscalar. Is there an analogue of $\alpha$ decay? In our world
this is due to two factors. The first is that there is an unique light
nucleus, $^4$He, which is extremely tightly bound. This is because two
nuclear shells close with $^4$He, and the next stable nucleus has mass
number 6. The second factor is that heavy nuclei become less tightly
bound as the baryon number increases. In \qcd, the shell structure sets
in at much higher values of $B$. As a result, there is no tightly bound
light nucleus which could be emitted to stabilize a heavy nucleus. Can
a heavy nucleus emit a single N to stabilize itself? This is possible,
since the first nucleus which has a nucleon in an upper shell may be
able to lower its energy significantly by ejecting a nucleon.  So a
rudimentary version of radioactivity is not ruled out.

This radioactive decay mode also opens up the possibility of spontaneous,
as well as induced, fission.  Such a nucleus with a $B$ just large enough
to send a few nucleons into a new shell could tunnel into a lower energy
configuration with two nuclei, both with all constituents in a lower
shell. Chain reactions are also possible.

Fusion is possible in spite of the absence of an unusually tightly
bound light nucleus like $^4$He. As long as the binding energy per
nucleon increases with $B$, it is possible to have a chain of nuclear
reactions which produce heavier nuclei. This chain would generally stop
when increasing $B$ by unity no longer increases the binding energy
per nucleon.

\subsection{Atomic physics and chemistry}

Let us move on to even lower energy. Uncharged atoms have
integer $Z$, increasing from unity, with baryon number, $B$, increasing in
steps of 2. There are no isotopes, so one always has $B=2Z$. There is no
analogue of the hydrogen atom of our world, with $B=1$ and $Z=1$. However
the atom with $Z=1$ and $B=2$, which we denote\footnote{We adapt the standard
notation for the nucleus of element X as ${}^B_ZX$, and drop
the index $Z$, since it is always $B/2$.}  as ${}^2H$ , has a
spectrum which is like that in our world, with a value of the Rydberg
constant which is the same as ours. In fact this sector of nuclei gives
rise to atomic physics which closely parallels that in our world,
except for a near absence of hyperfine structure, since nuclei have
vanishing spin (except when higher shells are occupied). This sector
of the periodic table is the same as in our world; and its chemistry
is nearly identical. The main difference is that the change in atomic
masses will change reaction rates somewhat.

On the other hand, there is also exotic atomic and molecular
physics. There is, for example, a charged bound state of N and $e$, which
is an exotic version of hydrogen with half integer negative charge,
${}^1H^-$. It has a spectrum similar to that of ${}^2H$ with energy
levels at half the value, so that the spectrum is shifted further to
the infrared.  Our world has a positively charged ion $H_2^+$, which is a
Hydrogen molecule with one electron removed. In the {\qcd} world one has
a molecule $({}^1H)_2^0$ which has one electron orbiting two separate
nuclei N. Positive and negative ions of different elements with half
integer charges are also possible. The change in the effective value
of the Rydberg in this sector changes reaction rates drastically. Redox
reactions with ${}^1H$ will proceed much faster than those with ${}^2H$,
for example.

Matter in the bulk could consist of uncharged molecules, either of the
kind which are familiar to us (like $({}^2H)_2$ or more complex molecules)
or the exotic molecules which we discussed (such as $({}^1H)_2$ or more
complex molecules) or molecules formed of familiar and exotic atoms.
(Could $({}^1H)_2{}^2H$ exist in this world?)  Electromagnetic plasmas are
also possible as bulk matter, as long as they are overall electrically
neutral. Either state of matter at the scale of $10^{23}$ molecules
will behave in a manner more or less familiar to us, albeit with detailed
properties which could be very different.

\subsection{Cosmology and astrophysics}

The physics of the inflationary and immediate post-inflationary epoch of
the universe is determined by physics beyond the SM, and we have
little understanding of it. Often this is dealt with by writing a low-energy
limit of this extended theory as an effective field theory built out of
only the SM fields. Much work has been done with this so-called SMEFT, and
many results could be taken over to this case. Since it is still an open
problem, we do not consider its effects on the imaginary universe we are
creating.

The physics of somewhat shorter cosmological length scales is determined
largely by the structure of the electroweak theory. One interesting question
here is the baryon asymmetry of the universe. Could the {\qcd} universe also
have a baryon asymmetry? In the SM $B-L$ (where $L$ is the lepton number)
is conserved due to the GIM mechanism and the consequent cancellation of
the ABJ anomaly. This generalizes to SM with any $N_c$ in the QCD sector.
Unless the SMEFT is radically different, one would expect that the mechanism
which gives rise to baryon asymmetry of our universe continues to operate
in this one\footnote{It is interesting that specifically for $N_c=2$ it is
easy to write terms in SMEFT which violate $B$ and $B-L$, through dimension 6
operators. This could allow a different path for baryon asymmetry to be
generated.}.

The history of this universe could diverge from that in ours a little
after the time of the {\qcd} phase transition. Primordial nucleosynthesis
(BBN) would certainly be different because of differences in the baryon
multiplets. For $N_f=3$ weak decays of the $su$ scalar baryon must
first give a $\Delta^{+++}$, which then decays into the N by emitting
a pion. This weak decay rate depends on the relative masses of the
scalar and vector baryons. So it may be possible to create hypernuclei
at BBN. Even otherwise the difference in the process of nuclear fusion
will change BBN.

The most interesting subsequent physics is the decoupling of photons
and matter.  In \qcd, this proceeds differently. The decoupling
temperature is a balance between the photon density, matter density,
and the Rydberg energy.  When there are sufficient number of photons
per atom in our universe to ionize a good fraction of hydrogen, one has
a plasma, and photons undergo Thompson scattering with ease. When the
temperature falls to a point that the probability of an atom being
ionized by a photon is small enough, then matter becomes largely
uncharged, and the sky becomes transparent. If the conditions at very
early times are similar, then the only difference between {\qcd} and our
world is in the atomic physics, assuming that we have tuned the nuclear
interactions so that the spectrum of $^2H$ in this world is similar to
our Deuteron. Primordial nucleosynthesis would produce a small amount of
${}^2H$, leaving most nucleons as they are. When the temperature falls to
3000 K, then ${}^2H$ atoms bind, but that still leaves a highly charged
plasma. When the temperature falls further, the remaining electrons will
bind into ${}^1H^-$, but there would still be a charged plasma. Since
the Thompson scattering cross section, $\sigma \propto e^2/m$, where
$e$ is the charge of the lightest charged species and $m$ its mass, the
mean free path of photons could increase substantially at this point,
but still remain much smaller than the Hubble radius. It is only when
the temperature falls sufficiently that neutral $({}^1H)_2$ molecules
form would the universe become transparent. Relic radiation in this world
would have a different temperature at a comparable age of the universe,
and the fluctuations in it could carry signatures of a time between the
formation of ${}^1H^-$ and decoupling.

The subsequent evolution of the universe involves gravity, and we know
no reason that at long-distances it knows about $N_c$. As a result, the
overall shape of the universe would remain the same. However, gas clouds
made of cold charged atoms will interact differently with radiation than
in our universe. The chemistry of gas clouds would be different. Their
collapse into galaxies and smaller objects is unlikely to change. Stars in
this universe would still be in self-regulating hydrostatic equilibrium
because there are analogues of nuclear fusion.  However the details of
stellar evolution will be very different since differences in nuclear
structure would change thermonuclear pathways radically. The flip side of
hydrostatic equilibrium is core collapse. Collapsed objects like black
holes, and analogues of white dwarfs and neutron stars are possible. So
although the stars, galaxies and gas clouds behave quite differently, the
overall appearance of the universe may not be very different from ours.

\section{Conclusions}

We examined the physics of a world where the standard model changes in
a single respect: the value of $N_c$ goes from 3 to 2. We minimized the
consequent changes in particle physics by insisting on retaining baryon
number conservation. In spite of this, a seemingly minor change at the
fermi scale induces a whole cascade of changes in the physics at long
distances. The changes in particle physics are expected.  We argued that
the physics of the phase diagram of \qcd, with baryon number conservation,
may differ from that in QCD only in minor (but interesting) ways. However,
at longer scales, the EFT which describes NN interactions changes.
At even longer length scales, nuclear and atomic physics, as well as
chemistry and astronomy, are significantly different.

This change in the physics at very long distances due to a single
change at short distances, may at first be surprising to a quantum
field theorist. On examining this in detail, one sees that it is due to
two facts: first that baryons are bosons, and second that baryons are
fractionally charged (in units of the electron charge). These are the
open windows through which UV physics escapes into the IR world. This is
therefore a cautionary tale about making too strong an assumption about
UV/IR decoupling in quantum field theory.  It is not only the global
symmetries that are important, but also the nature of the fields which
represent them.

From our discussions it is clear that fractionally charged bosonic
baryons are characteristic of all QCD-like theories with even $N_c$,
coupled to an unchanged remainder of the standard model. The long
distance physics of all these model worlds will differ from ours in the
ways that we outlined for $N_c=2$.  For odd $N_c$ baryons are fermions,
and carry integer charges. It is also possible to construct a nucleon
isodoublet of spin $1/2$ in all these models.  and arrange them to have
charges which are the same as in our world. These worlds will differ only
subtly from ours. The existence of galaxies and stars, stable nuclei and
complex chemistry means that the anthropic principle (in the sense used
in \cite{anthropic}) cannot be invoked to rule out any value of $N_c$
{\sl a priori\/}. In this context at least, if the anthropic principle
is used as a tool to decide on the viability of theories, then it is a
pretty blunt tool.

\begin{acknowledgements}
We acknowledge support of the Department of Atomic Energy, Government
of India, under Project Identification No. RTI 4002. We would like to
thank Jean-Paul Blaizot, Robert Shrock, and David Tong for helpful
comments.
\end{acknowledgements}



\end{document}